**To quote this work please use**

J.S. Wu, V. Brien, P. Brunet, C. Dong and J.M. Dubois
*Electron Microscopy study of scratch-induced surface microstructures in Al-Cu-Fe icosahedral quasicrystal*
Philosophical Magazine A, Vol 80, N°7, 1645-1655 (2000)
hal-02882489, doi.10.1080/01418610008212141

**Thanks**

# Electron microscopy study of scratch-induced surface microstructures in Al-Cu-Fe icosahedral quasicrystal

J.S. Wu[1,2], V. Brien[1], P. Brunet[1], C. Dong[3] and J.M. Dubois[1]

[1] Laboratoire de Science et Génie des Matériaux Métalliques (CNRS UMR 7548) and GRD CINQ,
Centre d'Ingénierie des Matériaux, Ecole des Mines - Parc de Saurupt, 54042 Nancy Cedex, France

[2] Beijing Laboratory of Electron Microscopy, Institute of Physics,
Chinese Academy of Sciences, P.O. Box 2724, 100080 Beijing, P.R. China

[3] Department of Materials Engineering, Dalian University of Technology,
116023 Dalian, P.R. China

Correspondence to:
Dr. J.M. Dubois
LSG2M, Ecole des Mines
Parc de Saurupt
54042 Nancy Cedex
France

Fax: +33-3 83 57 63 00
email: Jean-Marie.Dubois@mines.u-nancy.fr

**ABSTRACT**

Microstructure modifications induced by sliding a WC-Co indenter in scratch tests on the surface of a single phase AlCuFe icosahedral quasicrystal (IQC) was studied by scanning electron microscopy (SEM) and transmission electron microscopy (TEM). The scratch track was shown to comprise many smaller tracks. Dislocations were discovered to emerge from the edges of the smaller scratch tracks. Along a small track where shear stress is concentrated, a phase transition from IQC to a body-centered cubic (*b.c.c.*) phase with lattice parameter $a$=0.29 nm was pointed out. A modulated quasicrystal state as well as a deformation twin of IQC were determined in the region beneath the scratch.

**§1. INTRODUCTION**

Quasicrystals (QCs) are very brittle at room temperature, a drawback that limits the technological application of quasicrystalline alloys as bulk materials. However, using quasicrystalline alloys as coating materials on soft metals is promising since most of the QCs offer high hardness (Sainfort and Dubost 1988, Kang and Dubois 1992a) and low friction coefficients (Dubois, Kang and Stebut 1991, Wittmann, Urban, Schandl and Hornbogen 1991). For instance, soon after their discovery QCs have found a practical application as coating for cook-ware thanks to their significantly weak surface energy (Dubois and Weinland 1988, Dubois, Plaindoux, Belin-Ferré, Tamura and Sordelet 1998).

At high temperature, most of the QCs are plastically deformable starting from temperatures of roughly half that of their melting point (Shibuya, Hashimoto and Takeuchi 1990, Bresson and Gratias 1993). A dislocation glide mechanism was proposed to explain the plastic behaviour of the Al-Pd-Mn icosahedral quasicrystal since direct evidence was found for the creation of mobile dislocations during the deformation process (Wollgarten, Beyss, Urban, Liebertz and Köster 1993, Inoue, Yokoyama and Masumoto 1994). However, the deformation mechanism in the Al-Cu-Fe system appears more complex. Rather, Shield (1997) proposed a shear transformation similar to a martensitic transformation. Kang and Dubois (1992a) attributed the plasticity to thermal lattice vacancies and pore formation whereas Bresson *et al.* (1993) proposed essentially a phason-controlled deformation mechanism. Furthermore, deformation twining in icosahedral $Al_{65}Cu_{23}Fe_{12}$ strained by high-temperature creep was considered as another mode of plastic deformation (Shield and Kramer 1994). Perfection of the IQC structure could also affect the ductility and a highly defective IQC was shown to exhibit more ductility (Shield 1997).

The standard scratch test is often used to examine tribological properties (friction and wear) of materials and is extensively used in various research groups in order to test the potential application of QCs in contact mechanics. In these context, quasicrystal coatings show favourable tribological properties. For example, little but definite ductility was evidenced within the area sheared by the repeating passes of the loaded indenter whereas no ductility could be detected whatsoever outside the scratch tracks (Kang, Dubois and Stebut 1993, Stebut, Strobel and Dubois 1995). A detailed understanding of the deformation mechanism that takes place in the region beneath the surface of the specimen where the contact (or Hertz) pressure goes to a maximum should consider carefully the resulting microstructure. In icosahedral AlPdMn, Wollgarten and Saka (1997) have examined the microstructure of the deformed surface below Vickers indentations and proposed a deformation mechanism by grain boundary sliding. Since during the friction experiment, an elevated temperature can be reached within the zone in contact with the scratch indenter in dry sliding under high contact pressure, complex structural modifications caused by the combined effect of temperature and stress concentration are not unexpected. As stated by Stebut *et al*. (1995) stress-induced toughness enhancement in the wear track could originate from a stress-induced phase transformation. We address this question in the present study. The microstructure resulting from scratch testing a single-phase icosahedral $Al_{62}Cu_{25.5}Fe_{12.5}$ sample is investigated both by scanning and transmission electron microscopy. We show that a pressure induced phase transformation is indeed involved in the rise of ductility due to the severe scratch straining applied to the surface of the sample. A *b.c.c.* phase is detected within the core of the scratch track as well as a modulated icosahedral phase.

## §2. EXPERIMENTAL PROCEDURE

A disk (4 mm in height and 30 mm in diameter) of single phase, poly-grained $Al_{62}Cu_{25.5}Fe_{12.5}$ (*at*%) bulk specimen was prepared by sintering, starting from a powdered ingot of the icosahedral alloy. This alloy was produced by RF melting of the pure constituents under He atmosphere. The powder particle size was in the range 1 to 25 µm. Sintering was achieved under pure argon in a floating carbon matrix and external pressure of 30MPa. The temperature was rised first for 20 mn to 20K above the peritectic formation temperature of the icosahedral phase and then decreased to 20K below that reaction temperature. An annealing treatment was then performed at that temperature for 3 hours to make sure that no residual *CsCl*-type cubic phase could be retained in the sample. Slow cooling to room temperature at 20K/mn followed, yet maintaining the applied pressure. The sintered disk was then carefully mirror polished. A standard multi-pass and unidirectional scratch test was performed with a CSEM Revetest apparatus using a WC spherical indenter with tip diameter of 1.58 mm. The normal load was kept constant and equal to 30N. The pass number was 5 and the friction coefficient $\mu$=0.11. Initial specimen surface analysis was done by metallographic reflection light microscopy and with a Philips XL30 S FEG (field emission gun) scanning electron microscope operated at 5 kV with the best lateral resolution being 2.5 nm. Then the specimen was cut into small pieces having scratch tracks on their centre. Plane-view samples of those scratches were prepared by mechanical polishing employing a tripode thinning system that can reduce the specimen thickness to less than 1 µm and subsequent argon ion-milling at 3 kV. The Ar beam was directed to one side only of the thin sample in order to select the position of the final TEM specimen with respect to that of the bottom of the scratch. The diffraction contrast analyses were carried out in a Philips CM12 transmission electron microscope at an accelerating voltage of 120 kV. Micro-diffraction analysis was done in CM12 with a beam diameter of 80 nm.

## §3. RESULTS AND DISCUSSION

Figure 1(*a*) shows a SEM image of a scratch track produced on the surface of the IQC sample. The large white arrow indicates the indenter sliding direction. This direction is the same for fig.1(*b*) and (*c*). Partial ring cracks (labeled "*R*") have formed at irregularly spaced distances. This is typical of this dry sliding test. The width of the friction scratch is about 88 µm whereas those partial ring cracks may extend to almost 120 μm. It appears that the broad scratch track is comprising a large number of parallel thin tracks. More structural characteristics of these small tracks can be figured out at a larger magnification scale as shown in fig. 1(*b*). Small particles with a size of a few tens of nanometers, as indicated by small white arrows and labeled "*p*", are located along the scratch track. Instead of a flat surface, we see fluctuations of the surface contrast most plausibly due to the appearance of localized roughness indicated by "*F*" in fig. 1(*b*), implying that some plastic deformation has occurred. Small micrometric holes can also be found, possibly due to mechanical damage of hard particles located under the indenter during the test. In fig. 1(*c*), which was taken in the centre of the main scratch track, a partial ring crack propagating out almost perpendicular to the sliding direction is visible. Many particles with variable size (as signed by white arrows) can also be seen. It is known that a transfer layer, mainly consisting of re-deposited material, is formed between indenter and sample during scratch testing. Thus, most of the particles pointed out in fig. 1(*b*) and (*c*) must be re-deposited material. As indicated by small arrowheads in both fig. 1(*b*) and (*c*), a few mini-tracks of scratch can be found. Furthermore, some micro-cracks spreading out in a direction almost perpendicular to that of those mini-tracks (or the direction of indenter sliding) can also are observed. They are indexed by "*c*".

Figure 2(*a*) shows a TEM image of one of those mini-tracks passing through the multi-grained icosahedral sample. The large black arrow indicates the sliding direction of the indenter whereas the capital letters "*A-D*" label four different grains of AlCuFe IQC. No obvious decohesion or microstructure damage can be found along the grain boundaries. Rather, in grain B a dislocation (or possibly a micro-crack) is observed emerging from the track border, as indicated by a small white arrowhead. Although the process of strain release can occur during the thinning of the sample, black fringes caused by residual strain may still be observed. Along the scratch track, small particles with nanometer scale size can be discovered as shown in fig.2(*b*), which is a large magnification of the region indicated by a small white arrow in grain *D*, fig.2(*a*). Micro-diffraction was employed to determine the structure of those small particles as shown in fig.2(*c*) and (*d*). Figure 2(*c*) was taken by tilting the AlCuFe IQC matrix until one of its 2-fold directions came parallel to the electron beam whereas fig.2(*d*) was taken along one of its 5-fold axis. Except for the diffraction spots originating from the IQC matrix, the periodically spaced spots (indicated by small arrows in fig.2(*c*) and (*d*)) demonstrate that a *b.c.c.* phase or compound with lattice parameter *a*=0.29 nm has formed within the icosahedral lattice. The value of the lattice parameter is identical to that of the *β*-cubic, *CsCl*-type phase which occupied the middle region of the Al-Cu-Fe phase diagram (Faudot 1993). The spots due to the *b.c.c.* structure along its <111> and <101> directions are marked by asterisks in fig.2(*c*) and (*d*) and indexed. Note in fig. 2**(d)** that the (h00) spots, with h=±1, expected for the *CsCl*-type structure but not for a *b.c.c.* one are clearly missing. Hence, this *b.c.c.* cubic phase cannot be confused with any *CsCl*-type cubic phase formed in the specimen prior to deformation. The principal symmetry axes of IQC are also reported. The close orientation relationships between this cubic phase and Al-Cu-Fe IQC can thus be determined:

<110>; <113>  //  *A*5

and   <110>; <111>; <112>  //  *A*2.

The above orientation relationships are identical to those already found for co-existing AlCuFe IQC and its cubic phases (Zhang and Li 1990, Zhang, Feng, Williams and Kuo 1993, Wang, Yang and Wang 1993). Here, the cubic phase particles located within the scratch track have a fixed orientation relationship with the IQC matrix, thus implying that those particles can not be re-deposited material but are the result of a direct phase transformation from AlCuFe IQC. Since the final step of sample preparation was done by ion-milling, its influence should be taken into consideration. However, instead of a thin layer of crystalline structure covering the surface of IQC (Zhang, *et al*. 1993), we have here embedded particles coherent with the matrix lattice. To the best of our knowledge, this is the first time that such coherent cubic particles are observed within an icosahedral matrix. Furthermore, the effect of ion-milling was reduced by using a low working voltage, hence avoiding the appearance of additional diffraction spots in EDPs of IQC as was observed elsewhere (see e.g. fig.4 of Zhang *et al*. 1993). Kang and Dubois (1992b) have reported a pressure-induced phase transition in AlCuFe IQC with the discovery of a cubic phase ($a$=0.83 nm) and suggested that an irreversible transformation from quasicrystal towards a more simple phase was possible under compression. A confirmation that such a transformation may take place was supplied more recently by a detailed study of icosahedral Al-Li-Cu under uniaxial compression (Yu, Baluc, Staiger and Kleman 1995). Thus, we assume that the phase transition observed here was mainly due to the high shear strain achieved in the contact region with the indenter while temperature was most plausibly rather high owing to the poor thermal conductivity of both the WC-indenter and AlCuFe IQC. Indeed, a (too) crude estimate of the temperature increase within the contact area gives $\Delta T \cong 1000K$ if one assumes that the work of the tangential force due to friction is entirely transformed into heat. Although obviously overestimated, such a large increment of temperature may be responsible in itself for a transformation towards the disordered, high temperature state of the $\beta$-cubic phase.

Within a AlCuFe single grain containing a scratch track, electron diffraction patterns (EDPs) were taken from the area surrounding the track as shown in fig.3(*c*) and (*d*). As a comparison, witness EDPs taken on a thin foil prepared from the same AlCuFe sample but far from the scratched area are also displayed in fig. 3(**a**) and (**b**). For the sake of comparison, we carefully prepared all the thin foils according to the same thinning-down procedure. A close inspection of the pattern shown in fig.3(*c*) indicates that streaked satellite spots are found mainly in two sets situated along a 2-fold direction and a projection of a 3-fold direction (indicated by large white arrows in fig.3(*c*)) separated by 54° from each other. Owing to the appearance of satellite spots, one spot seems to extend to a set of triangle-shaped ones as indicated by a pair of perpendicular small arrows in the right-down part of fig.3 (*c*). Furthermore, several weak spots happen to be separated by equal interdistances as shown by three small white arrowheads in the right-top part of the EDP. The 2-fold EDP shown in fig.3 (*d*) was obtained by tilting the sample along the two-fold direction indicated by a white arrowhead in fig. *3(a)*. Sharp satellite spots can be observed along two 3-fold directions. As shown by small arrowheads, locally base-centred diffraction spots are now clearly visible, implying that a kind of long-distance ordering occurred in the AlCuFe IQC. A dark-field image (fig.3 (*e*)) obtained by selecting a set of triangle-shaped intense reflections indicated by "*d*" in fig.3 (*c*) shows a set of short striations in the right hand side half of fig.3 (*e*), confirming that a microstructure modulation has occurred in the icosahedral structure. A white large arrow in fig.3 (*e*) shows the direction of the scratch track, while a black small arrow indicates a dislocation emerging perpendicularly from the track. The crystallographic orientation of the grains separated by the dislocation line is slightly different and hence their imaging conditions are not identical. Such a typical intermediate stage, dubbed a modulated quasicrystalline structure, was pointed out by Audier, Bréchet, de Boissieu, Guyot, Janot and Dubois (1991) in the course of a reversible transformation from a rhombohedral crystalline approximant structure towards a perfect quasicrystal in $Al_{63.5}Fe_{12.5}Cu_{24}$. A similar structural modulation was also observed in the continuous decomposition of AlCuFe IQCs by Liu, Köster and Zaluska (1991). Comparing with the phase modulation activated by annealing, two characteristics of our observation should be noted. Firstly, no satellite spots can be observed along 5-fold directions in the 2-fold EDP as shown in fig.3 (*d*). Instead, satellites are situated along two 3-fold directions. Secondly, the fringes observed in the dark-field image (fig.3 (*e*)) are short and discontinuous. Since an icosahedral phase transition might be induced not only by phason and phonon defects, but also by a diffusion-driven composition modulation (Menguy, Audier and Guyot 1992), uncompleted atomic diffusion and complex phason state should be the reasons for the appearance of the modulated quasicrystal. This standpoint is in line with the highly out-of-equilibrium thermo-mechanical solicitation encountered during the very short duration of the scratch straining.

Beside the AlCuFe IQC modulation structure, a deformation twin was observed in the scratch-sheared area as shown in fig. 4. Figure 4 (*a*) and (*c*) show, respectively, the EDPs of each twin parts oriented along a 2-fold axis, while fig.4(*b*) is a composite pattern. In all three figures, some main directions of IQC are labeled and the typical rhombus-shaped strong diffraction spots are indicated by small arrows as well. The twinning relationship can be simply described as a rotation twin with 2-fold axis in each twin crystal separated by 72°. A detailed description of the crystallography of this kind of twin was given by Shield *et al.* (1994).

As well as griding (Boudard, de Boissieu, Simon, Berar and Doisneau 1996), sliding tests are expected to introduce mechanical damage and phason defects within the quasicrystal surface of tested samples. Figure 5 shows another thin specimen etched out of another grain of AlCuFe IQC with a scratch track passing through. A large particle labeled "*Q*" is identified as an IQC particle. It shows a 2-fold axis tilted by about 10° from that of the surrounding matrix. This later axis is perpendicular to the plane of the figure. New grain boundaries are created when smaller parts of IQC are broken out the matrix by the indenter. For all other particles of that kind, we found no simple orientation relationship whatsoever with respect to the matrix. From existence of the hole indicated by a large black arrow head, one can suspect that this region is near the surface since the sample was thinned from one side only. Except the mechanical damage of the IQC grains, no phase transformation can be detected along similar scratch tracks close to the surface of the sample. As discussed above, the phase transformation can only be found beneath the surface in the depth region where the contact Hertz pressure reaches its largest value (a calculation of the Hertz pressure with depth was reported for a QC by Stebut *et al.* (1995)). Again, grain micro-cracks (or dislocations) can be found as indicated by the small black arrows. It is interesting to note that at the end of dislocation labeled *D*, an array of small dislocations (labeled *A*) perpendicular to *D* is visible. A similar configuration of dislocation lines was discovered to stem from the crack tip during crack propagation in sintered alumina under micro-indention (Ikuhara, Suzuki and Kubo 1992). It seems that the observed dislocations act more like a micro-crack in grains of AlCuFe IQC and their mobility is still a question.

## §4. CONCLUSION

When produced under significantly high normal load on the surface of AlCuFe IQC, a scratch track was shown to consist of a lot of smaller scratch tracks associated with a network of emerging dislocations (or possibly micro-cracks). The existence of a transfer layer between indenter and sample, consisting mainly of re-deposited material, was confirmed. Splitting of IQC grains and generation of new grain boundaries were also observed. As a result of shear strain applied during the scratch test, a phase transformation from AlCuFe IQC to a *b.c.c.* structure with parameter *a*=0.29 nm was proved to occur in the region located below the surface where the contact Hertz pressure goes to a maximum (Stebut *et al.* 1995). A modulation of the quasicrystal structure was also found in the vicinity of the scratch. Although the multiplication of dislocations was observed in AlCuFe IQC, their mobility and contribution to deformation were uncertain. However, the enhancement of ductility which goes along with repeated indenter sliding might result from a stress-induced phase transition, deformation twinning and the appearance of a defective quasicrystal phase. Due to the strong coupling between electron localization on the one hand and influence of defects or of impurity phases with metallic character on the other hand, the present study has direct relevance to the control of quasicrystal surfaces subject to mechanical polishing or grinding in view of certain technological applications.

**ACKNOWLEDGEMENTS**

This research was financed by the Chinese-French Advanced Research Program (PRA MX 96-02). We acknowledge the financial support offered by CNRS, Région Lorraine and CUG-Nancy (CPER 1994-99) and by the National Natural Science Foundation of China (grant No. 59525103). We thank J. von Stebut for the provision of scratch-test facility and him and C. Comte for interesting discussions. We are most grateful to G. Beck and K.H. Kuo for their constant interest in this work.

**Figures**

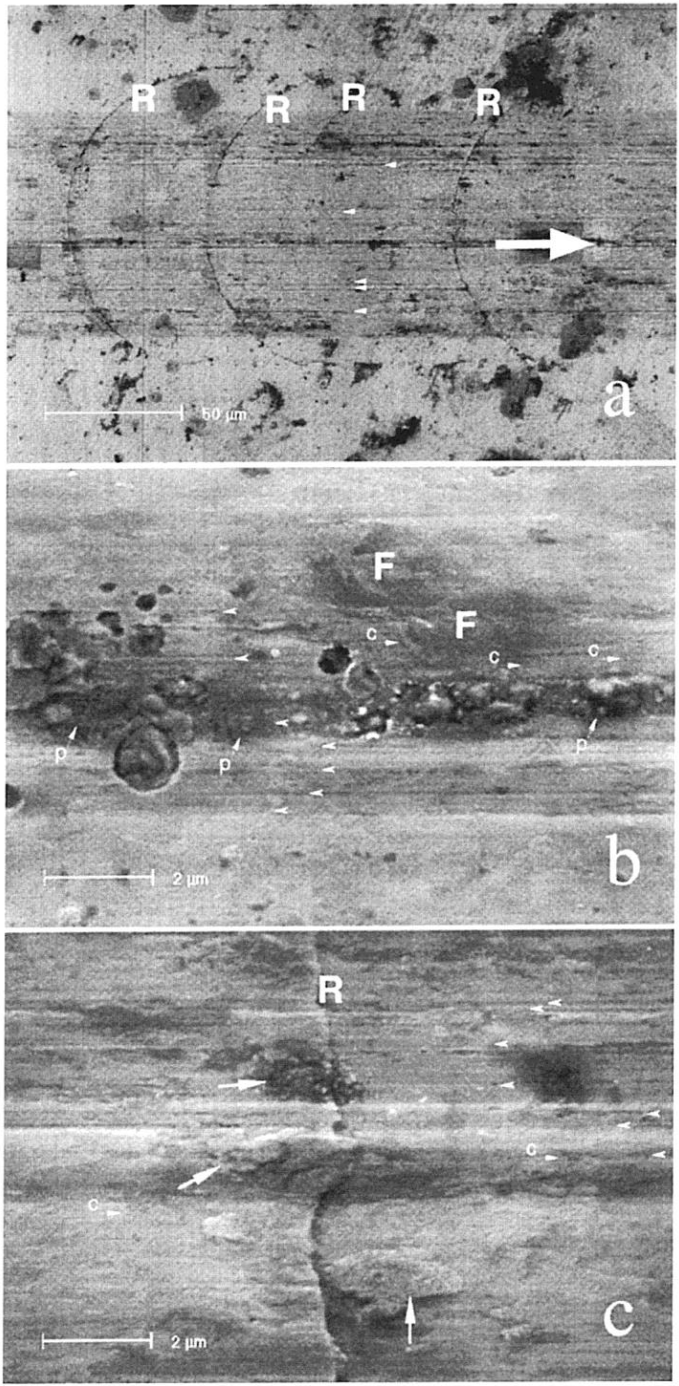

Fig.1: (*a*) SEM image of a scratch track produced by a WC indenter at 30 N of load on the surface of a AlCuFe IQC single phase sample; (*b*) and (*c*) are larger magnifications of a small track and of a partial ring crack, with small particles "*p*", micro-cracks "*c*", contrast fluctuations of the surface "*F*" and traces of mini-scratch labeled with small arrow head.

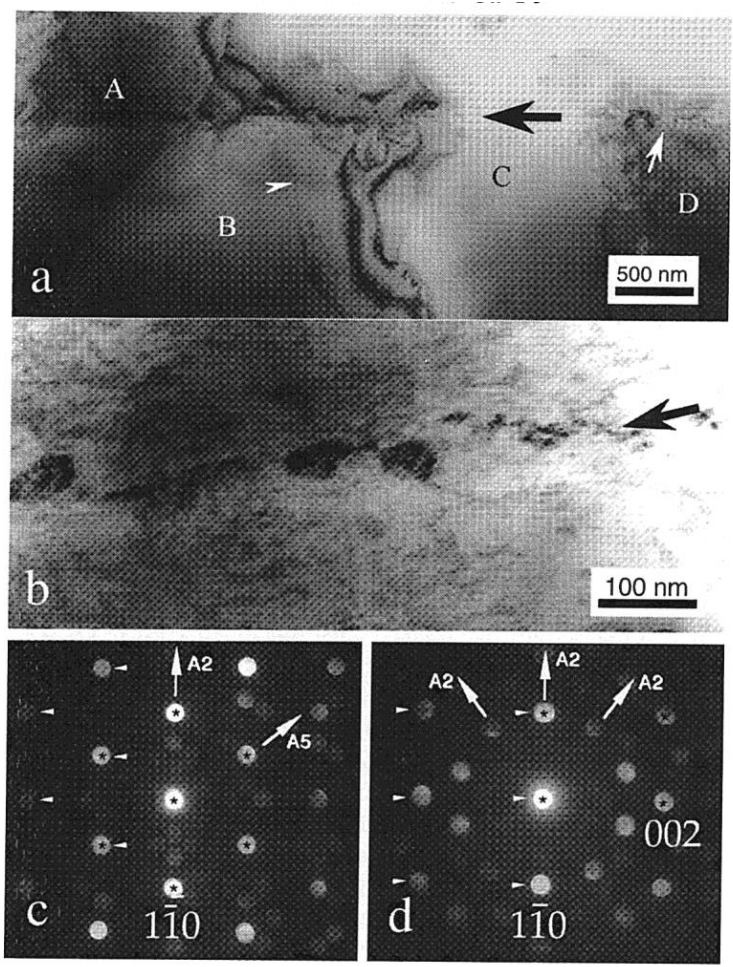

Fig.2: (*a*) scratch track propagating through four grains (A, B, C and D) of AlCuFe IQC, (*b*) nanometer scale particles with a *b.c.c.* structure (*a*=0.29 nm) formed along the scratch track, (*c*) micro-diffraction pattern of one of those particles in (*b*) when the orientation of the AlCuFe IQC matrix is along a 2-fold axis and (*d*) when the matrix is oriented along a 5-fold direction.

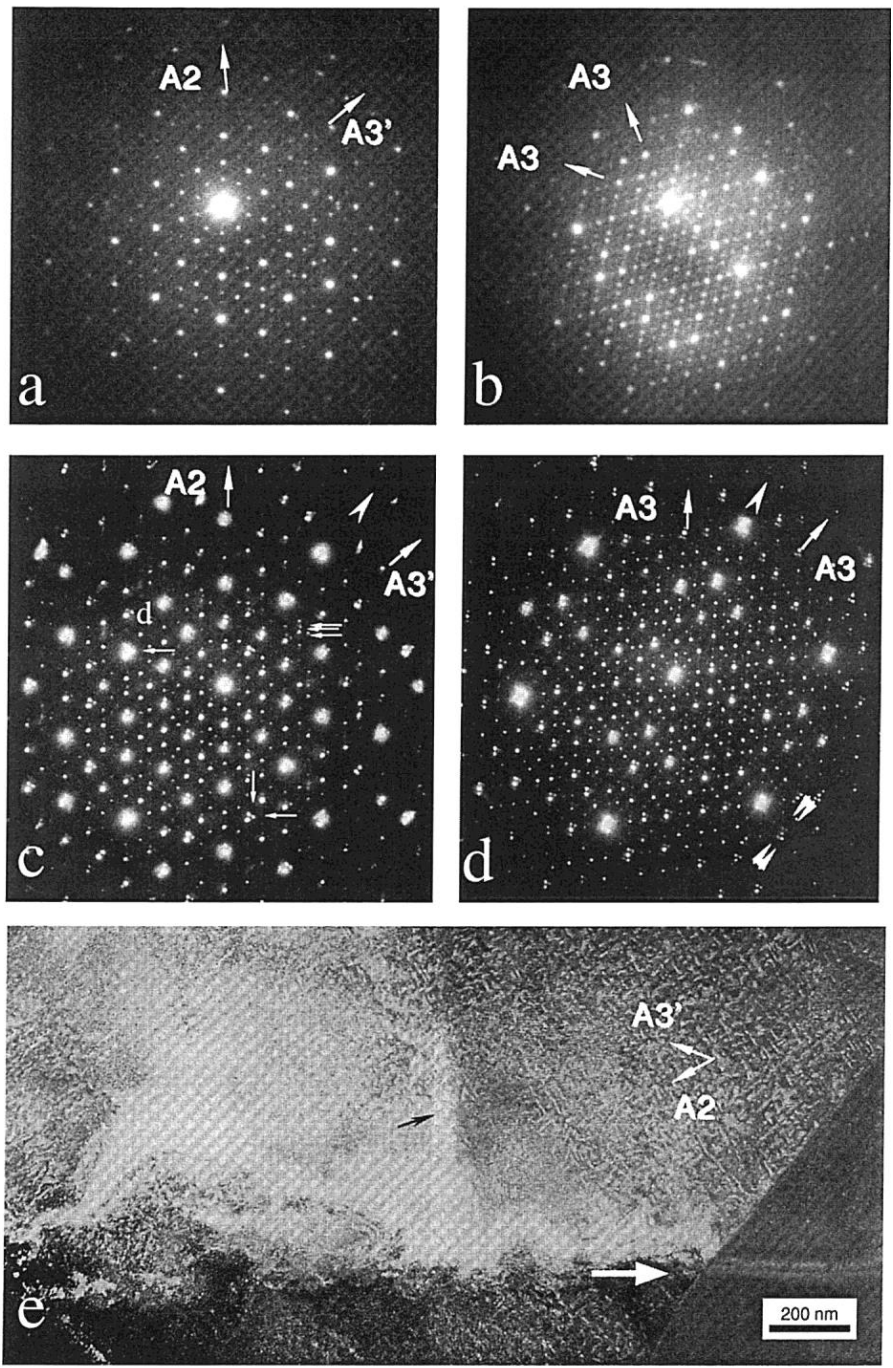

Fig.3: Modulation of the AlCuFe IQC provoked by the scratch test. In (*a*) and (*b*), reference EDPs of the AlCuFe IQC phase prior to scratch straining are shown taken along a 5-fold axis (*a*) and along a 2-fold axis (*b*). Electron diffraction and imaging performed in the vicinity of a micro-scratch : (*c*) along a 5-fold axis and (*d*) along a 2-fold axis, superimposing with satellite spots; (*e*) dark-field image obtained by selecting a set of triangle-shaped intense reflections marked *d* in (*c*). Note the short and discontinuous striations. The black arrow indicates the position of a dislocation.

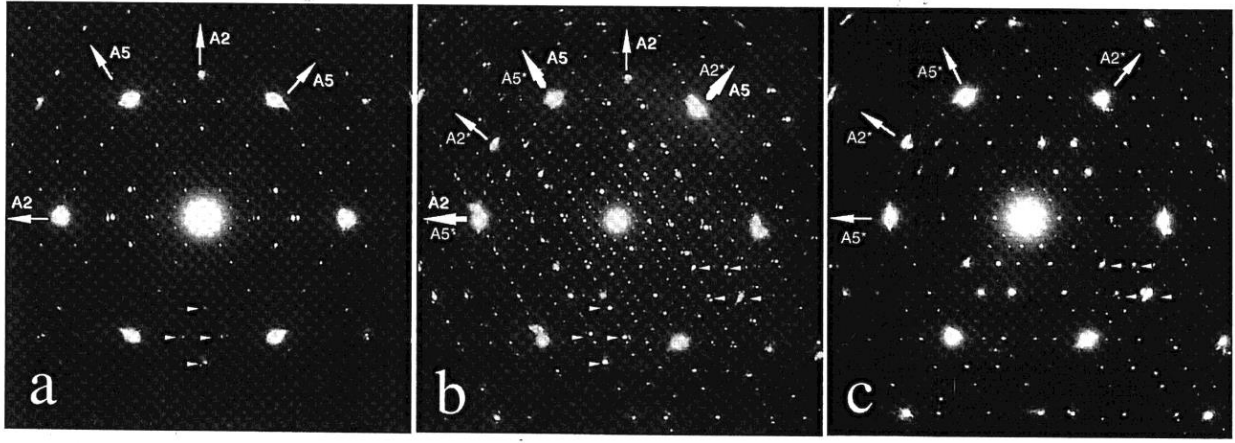

Fig.4: EDPs of (*a*) one twining part, (*b*) composite pattern and (*c*) another twining part of a deformation twin of AlCuFe IQC oriented along a 2-fold axis in the region of the scratch track. A few main directions of IQC are indicated with arrows and typical rhombus-shaped strong diffraction spots by small arrow head.

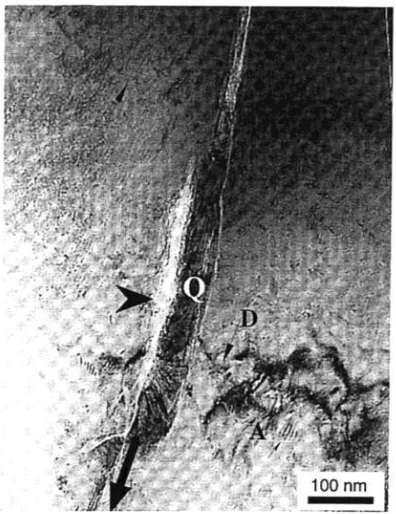

Fig.5: TEM image of a grain of AlCuFe IQC containing a scratch trace (small black arrow). A small fraction "*Q*" of the grain was cut down by the indenter, leaving open the grain boundary labeled by the black arrow head. Note the array "*A*" of dislocations formed ahead the main dislocation line "*D*".

**To quote this work please use**
J.S. Wu, V. Brien, P. Brunet, C. Dong and J.M. Dubois
*Electron Microscopy study of scratch-induced surface microstructures in Al-Cu-Fe icosahedral quasicrystal*
Philosophical Magazine A, Vol 80, N°7, 1645-1655 (2000)
hal-02882489, doi.10.1080/01418610008212141
**Thanks**